\documentclass{elsart}   

\usepackage{graphicx}
\usepackage{epsfig}

\newcommand \be{\begin{equation}}
\newcommand \bea{\begin{eqnarray}}
\newcommand \ee{\end{equation}}
\newcommand \eea{\end{eqnarray}}

\begin{document}

\begin{frontmatter}

\title{Inverse Statistics in the Foreign Exchange Market}
\author{M.H.Jensen,\corauthref{auth1}}
\ead{mhjensen@nbi.dk}
\author{A.Johansen,\corauthref{auth2}}
\ead{anders-johansen@get2net.dk}
\author{F.Petroni,\corauthref{auth3}}
\ead{fil\_petroni@tin.it}
\author{I.Simonsen\corauthref{auth4}}
\ead{Ingve.Simonsen@phys.ntnu.no}
\address[auth1]{Niels Bohr Institute, Blegdamsvej 17, DK-2100 Copenhagen {\O},
Denmark}
\address[auth2]{Teglg\aa rdsvej 119, DK-3050 Humleb\ae k, Denmark}
\address[auth3]{Dipartimento di Matematica and 
                 I.N.F.M. Universit\`a dell'Aquila, I-67010 L'Aquila, Italy}
 \address[auth4]{Department of Physics, NTNU, NO-7491 Trondheim, Norway}

\date{\today}
\maketitle


\begin{abstract}
  We investigate intra-day foreign exchange (FX) time series using the
  inverse statistic analysis developed in~\cite{hori1,hori2}. Specifically,
  we study the time-averaged distributions of waiting times needed to
  obtain a certain increase (decrease) $\rho$ in the price of an investment.
  The analysis is performed for the Deutsch Mark (DM) against the \$US for
  the full year of 1998, but similar results are obtained for the Japanese
  Yen against the \$US. With high statistical significance, the presence
  of ``resonance peaks'' in the waiting time distributions is established.
  Such peaks are a consequence of the trading habits of the markets
  participants as they are not present in the corresponding tick (business)
  waiting time distributions. Furthermore, a new {\em stylized fact}, is
  observed for the waiting time distribution in the form of a power law Pdf.
  This result is achieved by rescaling of the physical waiting time by
  the corresponding tick time thereby
  partially removing scale dependent features of the market activity.
\end{abstract}


\end{frontmatter}


\section{Introduction}

Per Bak was a great scientist and a fantastic source of inspiration
for many of us over many years. Through numerous lively and exciting
discussions with him, one always felt that a project or a calculation
was brought back on track again by his clever comments and
suggestions. He applied his ingenious idea of ``Self-Organized
Criticality'' to many different systems ranging from sand piles,
earthquakes to the brain and even finance. As he said: ``It's all the
same'', meaning that in the end the paradigm of the sand pile model
would after all describe the behavior of the particular system he
considered. The idea of applying inverse statistics to turbulence data
was the subject of the discussion between Per Bak and one of us (MHJ)
several times. He liked the idea, and as such, we are quite sure that
he would have liked our application of inverse statistics to financial
data. This in particular applies to the scale invariant power-law
scaling that is being observed for the normalized waiting time
distribution.  It is therefore our pleasure to dedicate this paper to
his memory.

With the financial industry becoming fully computerized, the amount of
recor\-ded data, from daily close all the way down to tick-to-tick level,
has exploded. Nowadays, such tick-to-tick high-frequency data are readily
available for practitioners and researchers alike. In general, such
high-frequency data are irregularly spaced in (physical) time, since
an actual trade is a negotiation between sellers and buyers through a
bid and ask process highly influenced by the irregular flow of information
reaching the market.  Hence, in order to apply the classic return approach
to such data, the asset price has to be re-sampled equidistantly in physical
time. This has been suggested in the seminal paper on high-frequency
foreign exchange data analysis published by the Olsen \& Associates
Research Institute~\cite{Olsen}, but in many ways such a re-sampling
violates the true dynamics of the market. Consequently, there has been
an increasing interest over the past decade in studying variations in
the market over a {\it variable} time span opposed to that of a {\it fixed}
time span as for the return distribution~\cite{hori1,hori2,wait}. One such approach
is to consider drawdowns/ups, where an increasing or decreasing trend is
followed to the end~\cite{outliers1,outliers2}. Recently, the present authors MHJ,
AJ and IS introduced another such time varying approach --- the
{\em inverse statistics approach}~\cite{hori1,hori2}.  At the heart of this
technique, lies the waiting time needed to cross a pre-described return
barrier\footnote{One may also consider the completion process of a trade
as the crossing of the bid and ask random walks.}. The distribution of
these waiting times, also termed investment horizons, characterize the
inverse statistics~\cite{Mogens} and has successfully been applied to
daily close stock index data~\cite{hori1,hori2}.

The purpose of the present paper is to follow up on these studies and
investigate the corresponding statistical distributions for the
foreign exchange (FX) market using high-frequency data. In particular,
this work focuses on the exchange rate for the full year of
1998 between the two major currencies of the world, namely the \$US
and the DM, the latter in 2000 replaced by the Euro.

\section{Formalism}
Before we present the results of our analysis, we will set the stage
by recapitulating a few important definitions and properties of
inverse statistics. A more detailed introduction can be found in
Refs.~\cite{hori1,hori2,Mogens}. Let us assume the value of the asset under
study is described by the time varying asset price $S(t)$\footnote{As
  the true trading price is not publicly disclosed, we have chosen to
  calculate the price as $S(t)=(S_{bid}(t)+S_{ask}(t))/2$. Other
  options are to use $ s(t)=(\log S_{bid}(t)+\log S_{ask}(t))/2$
  suggested in Ref.  \cite{Olsen} or the algorithm proposed in Ref.
  \cite{Filippo}.}.  Here, the time variable $t$ can in principle be
any time variable and below we will use both physical and tick time.
The log-return at time $t$ calculated over a time interval $\Delta t$,
is defined as
\begin{eqnarray}
  \label{Return}
  r_{\Delta t}(t) &=& s(t+\Delta t)- s(t),
\end{eqnarray}
where $s(t) = \ln S(t)$.  The waiting time for an investment made at time $t$
at log-price $s(t)$, is defined as the time interval $\Delta t = t'-t$, $t'>t$,
where the relation $r_{\Delta t}(t)\geq \rho$ is fulfilled for the {\it first}
time. If physical time is used as the time scale, then the waiting time for
return level $\rho$ is denoted by $\tau_\rho(t)$. If tick time is used instead,
the corresponding  (dimensionless) waiting time is denoted by $T_\rho$.
The investment horizon, or waiting time distributions are the
probability density functions of $\tau_\rho(t)$ and $T_\rho$ when
using physical or tick time, respectively.

For a geometrical Brownian motion this distribution, known as the {\em first
passage distribution}, is known analytically~\cite{Book:Karlin1966,DR1,DR2} to be
$p(t) = a\exp\left( -a^2/t\right)/\sqrt{\pi}\, t^{3/2}$, where $a$ depends on
the return barrier $\rho$. In~\cite{hori1,hori2} it was shown that this distribution
is too ``primitive'' to fit the waiting time distributions for the three major
US stock market indexes (DJIA, SP500 and NASDAQ) and instead a type of
generalized Gamma distribution was found to give an excellent parameterization
of the data, see Refs.~\cite{hori1,hori2} for details. As the daily close of the stock
indexes was analyzed, which by definition are regularly sampled with the
exception of weekends and public holidays, the distinction between physical
time and ``tick-time'' was not made.

%
\section{Analysis of the FX Market}
%

We have been able to obtain foreign exchange~(FX) data ({\it cf.}
Ref.~\cite{Olsen} for ``stylized facts'' of FX-markets) for the DM
against the \$US for the full year of 1998. The data set consists of
$1,620,944$ ticks irregularly distributed in physical time. This
corresponds to an average time between ticks of roughly $20$ seconds.
However, as we will see, there are hours during the day where the
trading activity is much higher than during the remaining of the 24
hour day. As we shall see, this will play an important role for our
results.

\begin{figure}[tbhp]
  \centering
    \includegraphics*[width=10cm,height=7cm]{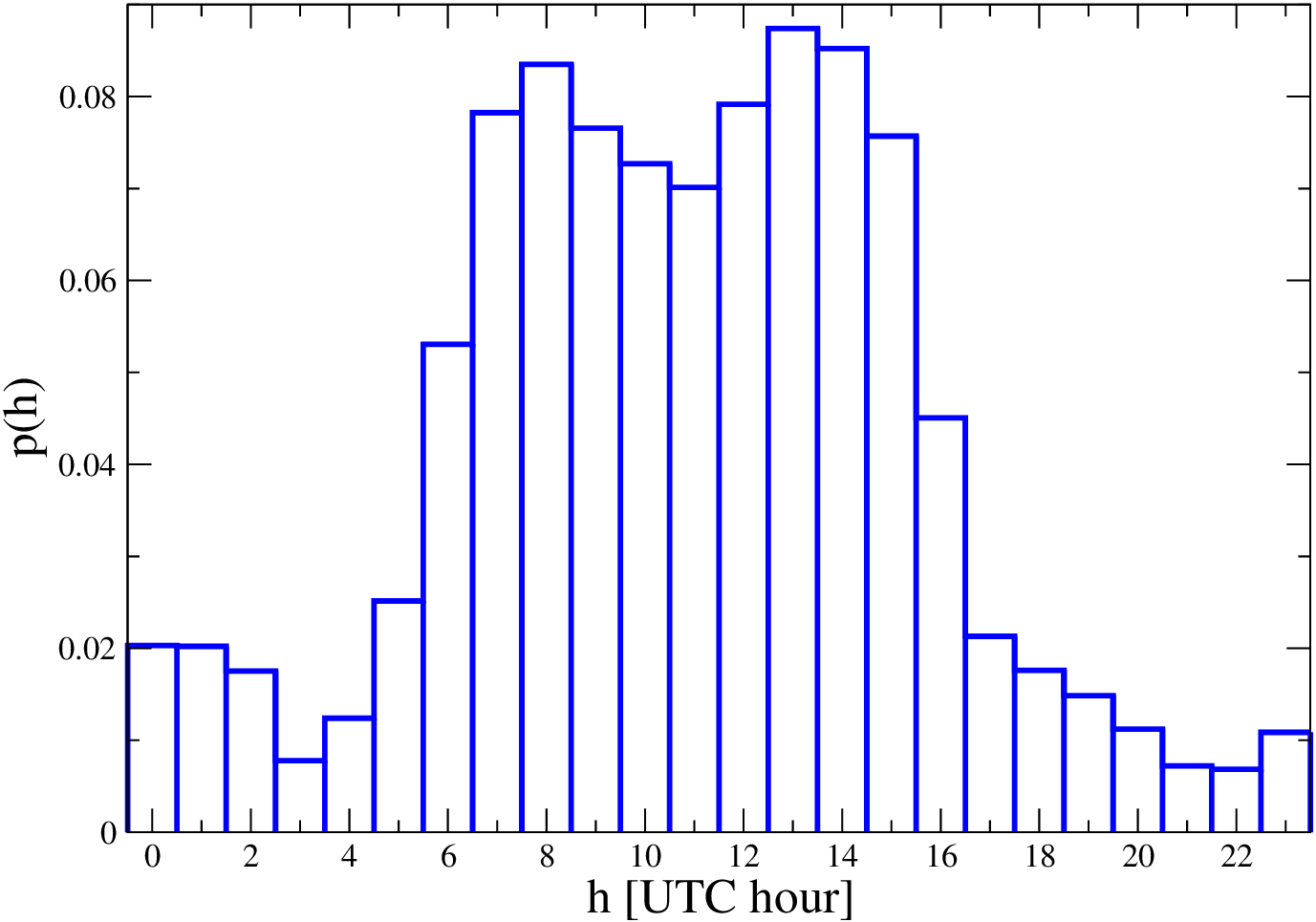}
  \caption{The tick frequency probability distribution,
    $p(h)$, {\it vs.} UTC-hour, $h$, at which the tick took place for
    the DM/US\$ of the year 1998.}
  \label{Fig:DMUS98-UTC-Hours}
\end{figure}

For high-frequency data, the results depend highly on how ``time'' is
defined~\cite{Olsen,Filippo}. Two obvious choices for a time scale are
{\em physical time} (or ``wall time'' displayed on the trading floor)
and {\em tick time} (also referred to as ``business time'' by some
authors) as mentioned previously. As we see in Fig.
\ref{Fig:DMUS98-UTC-Hours} the average physical time interval between
ticks will decrease during active market periods and on the other hand
increase when the market comes less active.


\begin{figure}[thbp]
  \centering
    \includegraphics*[width=10cm,height=7cm]{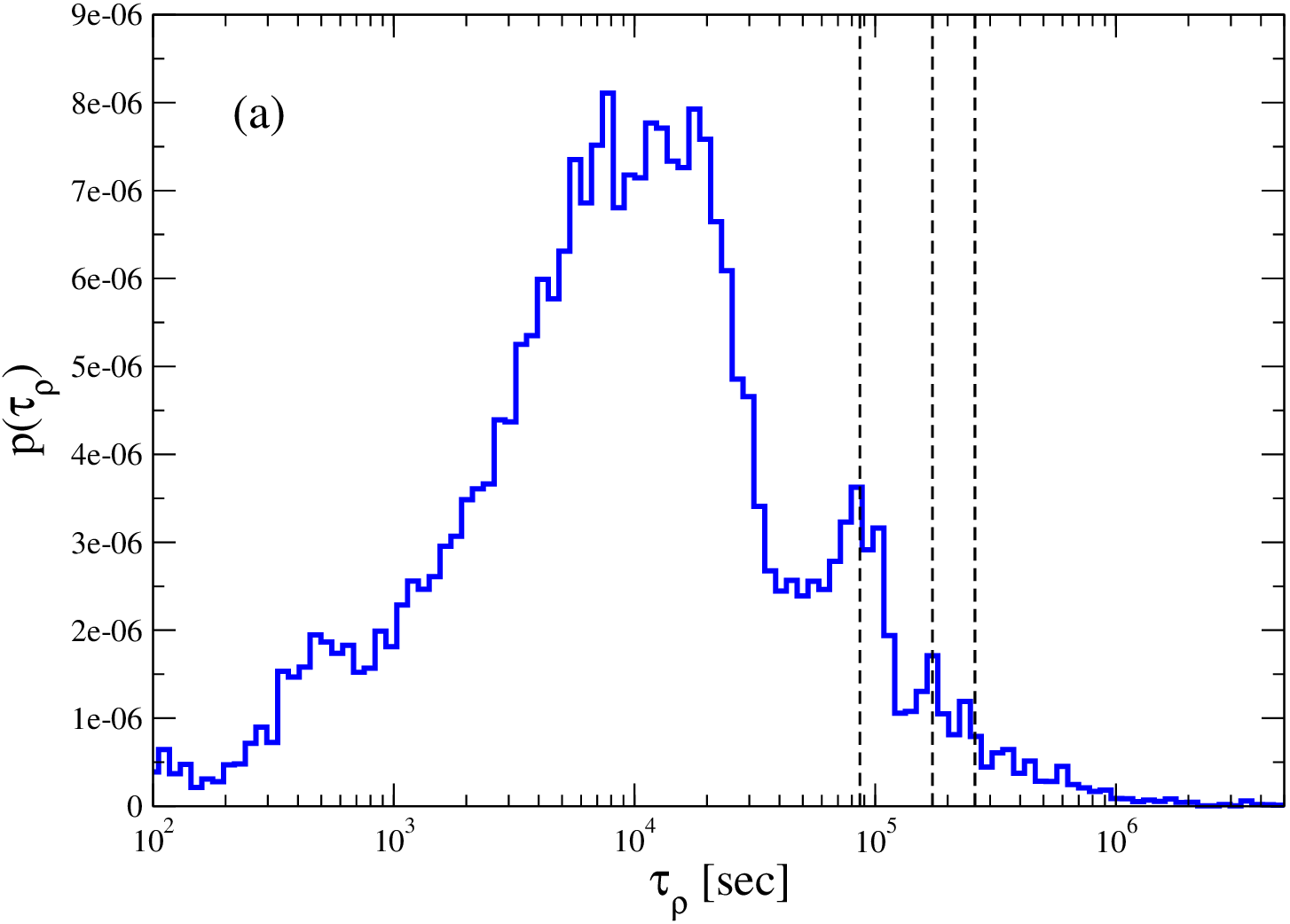} \\
    \includegraphics*[width=10cm,height=7cm]{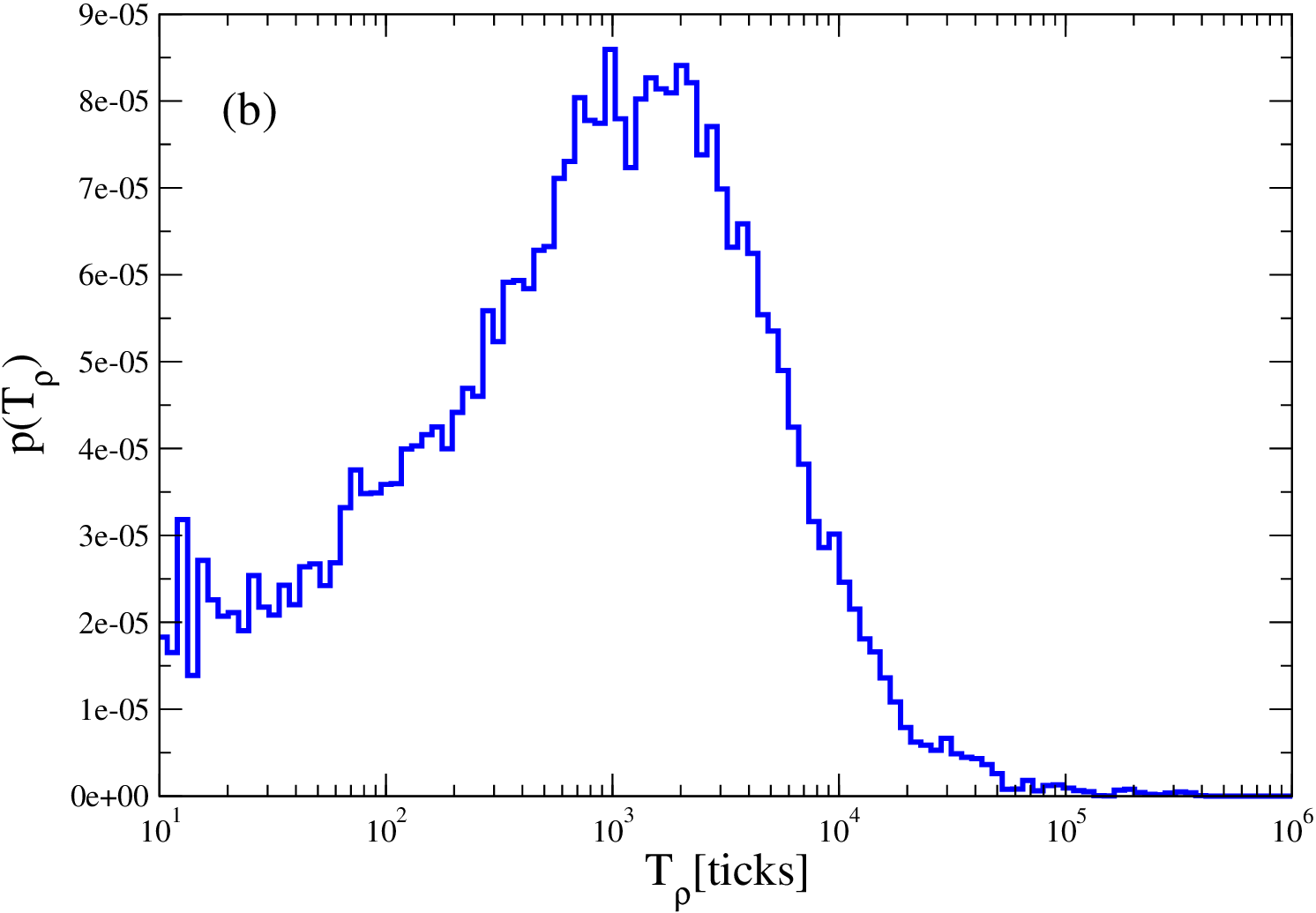}
  \caption{Inverse statistics ($p(\tau_\rho)$ or $p(T_\rho)$)
    {\em vs.} waiting time using {\em physical waiting time}
    $\tau_\rho$ (a) or {\em tick time} $T_\rho$ (b) for the 1998
    DM/US\$ data. The return level used to obtain these results was
    $\rho=0.005$. The vertical dashed lines in
    Fig.~\protect\ref{Fig:DMUS98-WatingTime}(a) indicate physical
    waiting times of $1$ to $3$ days (from left to right).  Notice the
    apparent resonances seem to coincide with the daily structure,
    while such resonances are not present in the corresponding tick
    time distribution.}
  \label{Fig:DMUS98-WatingTime}
\end{figure}

In Figs. \ref{Fig:DMUS98-WatingTime} a and b, the {\em physical waiting time}
distribution $p(\tau_\rho)$ and the {\em tick waiting time} distribution
$p(T_\rho)$ for the DM against the \$US is shown for the year 1998. The return
level used to obtain these results was $\rho=0.005$. We did also check that our
findings were not affected in any significant way by instead considering a
return level of $\rho=-0.005$. This indicates that drift is not an important
component to the analyzed data set, as opposed to the daily data analyzed
in~\cite{hori1,hori2}, and hence no need for detrending is present. The two figures
of~\ref{Fig:DMUS98-WatingTime} both go through a single maximum and for waiting
times smaller then these maxima the two distributions are similar. However,
for longer waiting times, there are some notable differences between the two
distributions. First, the tick time distribution, $p(T_\rho)$, falls off
faster, actually as $1/T$, than the corresponding distribution using
physical time. Secondly, and more important, the ``resonance peak''
structure present in  $p(\tau_\rho)$ has vanished in $p(T_\rho)$. The first
of these peaks is located roughly at the daily scale (indicated by the left
vertical dashed line in Fig:~\ref{Fig:DMUS98-WatingTime}(a)) and we clearly
see the second and third ``harmonics''.

The origin of these ``resonances'' is to be found in the varying activity
of the market. The main difference between the two ways of quantifying time
is that tick time is equidistant, whereas physical time between ticks is not.
Mathematically, one may say that the data using physical time is the
``convolution'' of the data using tick-time with the distribution of ticks
as a function of physical time. Hence, a change in market activity will alter
the inter-relation between these two time scales.  In order to study the daily
peak structure of Fig.~\ref{Fig:DMUS98-WatingTime}(a) in more detail, we in
Fig.~\ref{Fig:DMUS98-UTC-Hours} show the Pdf of ticks as a function of the
UTC (Universal Time Coordinate, former GMT) hour of that tick. One observes
that this distribution is far from being flat. Thus there exists indeed time
periods where the FX-market is semi-closed. In particular, almost $80\%$ of
the ticks correspond to a UTC-hour of $6$ to $16$ with a local maximum located
around UTC-hour $8$ and another one at $13$ or $14$.  The active periods
defined by UTC-hours from $6$ to $16$, correspond to working hours in London
and the east coast and the mid-west of the US.


In view of the results of Figs.~\ref{Fig:DMUS98-WatingTime} and
\ref{Fig:DMUS98-UTC-Hours}, one might suspect that the daily peak
structure observed in $p(\tau_\rho)$ is a result of this uneven
trading activity during the day of the global FX-market. If it is
(partly) true that (tick time) returns calculated from two consecutive
ticks are only weakly correlated, one would naively expect that the
volatility of a given {\em physical time interval} is larger in a high
market activity period than in a low one. Under this assumption, the
tick time distribution $p(T_\rho)$ should not be sensitive to whether
or not one is in a high or low activity region, since tick time by
construction is equidistant.  On the other hand, for the physical
waiting time distribution, $p(\tau_\rho)$, the market activity does
indeed matter. Here the pre-described return level will more likely be
reached during the highly active periods. If the return level is not
reached within one and the same period of high activity, there is
higher chance that it will do so in the next one than in the
intermediate low activity period, simply because there are fewer ticks
during this low activity periods. Such a behavior will therefore
result in an enhanced physical waiting time probability corresponding
to an integer numbers of days, just like we see in
Fig.~\ref{Fig:DMUS98-WatingTime}(a).

To investigate this further, we introduce a new type of time scale,
specifically a normalized waiting time that aims at partly suppressing
the effect of varying market activity.  This time scale is defined as
$\tau_\rho/T_\rho$, or in words, as the average {\em physical} waiting
time per tick needed to break through the return level $\rho$. As we
will see, normalizing the physical waiting time with the
(corresponding) number of ticks needed to cross the $\rho$ return
barrier, reduces the effect of varying market activity.  It should
also be noted that one naively would expect the inverse statistics, as
characterized by the normalized waiting time distribution, to be less
sensitive to the level of return $\rho$ then their unnormalized
partners. This is so since an increase in $\left|\rho\right|$ will
increase the overall waiting time measured both in physical or tick
time units.


\begin{figure}[tbhp]
  \centering
    \includegraphics*[width=10cm,height=8cm]{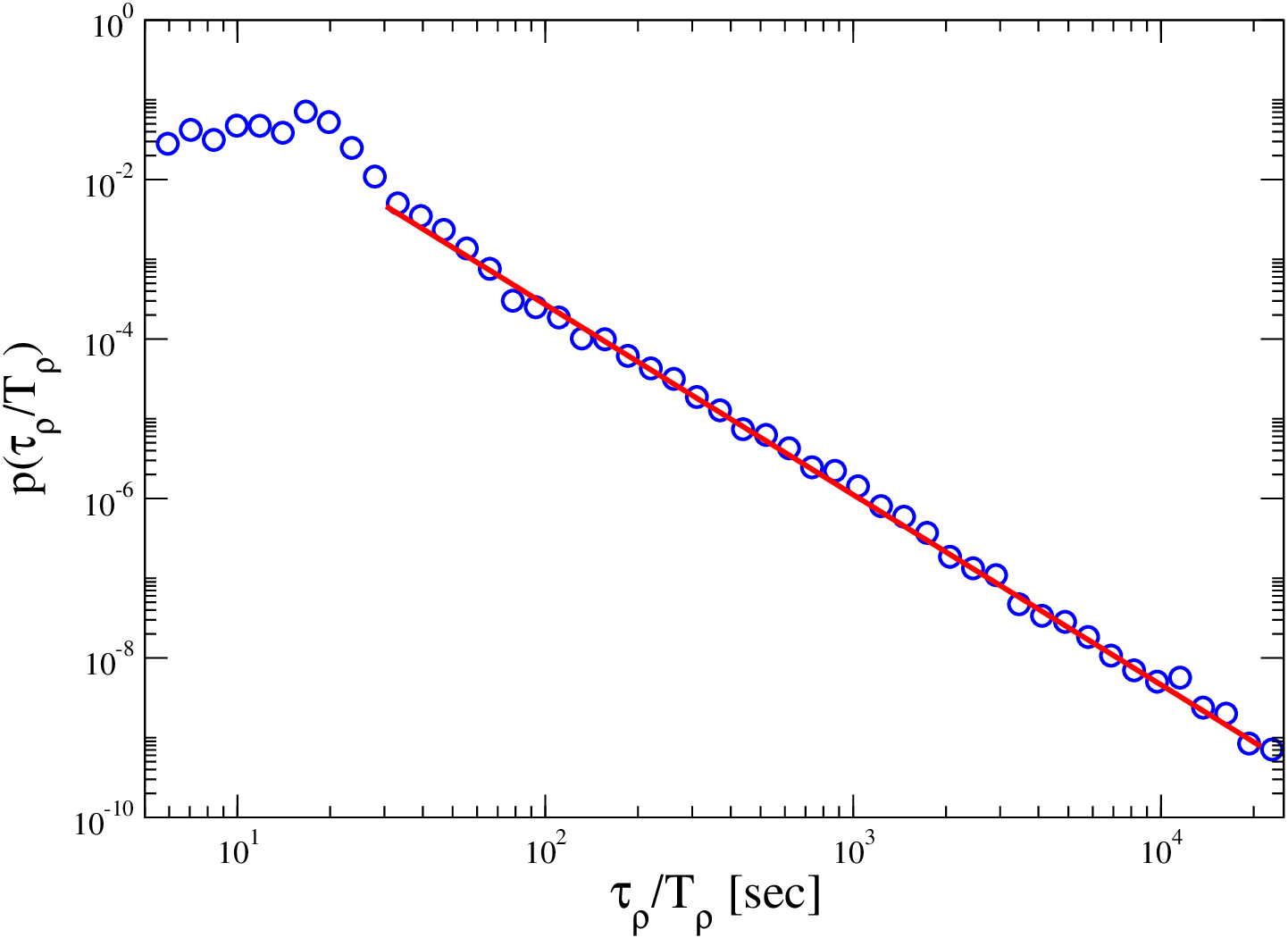}
  \caption{The probability distribution of {\em normalized} waiting
    time $\tau_\rho/T_\rho$ needed to reach a return level $\rho=0.005$
    for the 1998 DM against the \$US  exchange rate data
    (open circles).  The power law dependence $p(\tau_\rho/T_\rho)\sim
    \left(\tau_\rho/T_\rho\right)^{-\gamma}$ with $\gamma\simeq 2.4$
    is indicated by the solid line. }
  \label{Fig:DMUS98-Normalized-WatingTime}
\end{figure}


In Fig.~\ref{Fig:DMUS98-Normalized-WatingTime} the probability
distribution function for the normalized waiting time,
$\tau_\rho/T_\rho$, is presented. As suspected, there seems to be
little, or no, effect of the change in market activity throughout the
day. For instance the daily peaks that are so marked features of
$p(\tau_\rho)$ are now hardly observable in
Fig.~\ref{Fig:DMUS98-Normalized-WatingTime}.  However, more
surprisingly, the behavior of $p(\tau_\rho/T_\rho)$ for not too low
normalized waiting times, seems to be well fitted by a single
power-law. In particular one finds
\begin{eqnarray}
  p\left(\frac{\tau_\rho}{T_\rho}\right)   &\sim&
         \left(\frac{\tau_\rho}{T_\rho}\right)^{-\gamma},
         \label{Eq:Inverse-Stat-Normalized-Wating-time}
\end{eqnarray}
with $\gamma \simeq 2.4$ when $\rho=0.005$, spanning nearly three
orders of magnitude in normalized time $\tau_\rho/T_\rho$. (The
question of how sensitive $\gamma$ is to the return level $\rho$ will
be addressed in a separate forthcoming publication.) The conclusion
is that the proposed rescaled time is the most natural one to use when
analy\-zing high-frequency data in terms of inverse statistics and makes
the inverse statistics approach well suited for high-frequency  data.

\section{Conclusions}

In conclusion, we have studies high-frequency foreign exchange data
for the DM against the \$US  from an inverse statistics
point of view.  It is found that the change in market activity makes
it more challenging to define an appropriate and unique time scale,
since the change in activity level of the market causes certain
resonances to emerge in some quantities. In particular, when physical
time is used as the time scale it is demonstrated that daily peaks
emerge in the inverse statistics as quantified by the {\em physical
  waiting time} distribution function. Such peaks are, however, not
fond to be present in the corresponding inverse statistics for {\em
  tick time}. The trading activity effect is partly removed from the
inverse statistics by studying the new time scale defined as the
average physical waiting time per tick, $\tau_\rho/T_\rho$ needed to
reach a given level of return $\rho$. In terms of this normalized time
variable a new type of power law is observed for the inverse
statistics. Over a nearly three orders of magnitude in normalized
waiting times, excluding the smallest ones, the waiting time
distribution for $\rho=0.005$ was found to be well characterized by a
single power law of exponent $\gamma\simeq-2.4$. This scaling law
represents a new type of {\em stylized fact} for the FX-market which, to
the best of our knowledge, has not been reported before.

\section*{Acknowledgement}
The authors are grateful to M. Serva for providing
the data analyzed in this paper.


\end{document}